\documentclass[11pt]{article}
\usepackage{graphicx}
\usepackage{amsmath,amssymb}
\usepackage[latin2]{inputenc}

  \def\be{\begin{equation}}
\def\ee{\end{equation}}  
\def\ba{\begin{array}{c}}

\def\ea{\end{array}} 
\def\bea{\begin{eqnarray}}
\def\eea{\end{eqnarray}}
 
\begin{document}

\begin{center}
{\Large\bf Trigonometric identities, angular Schr\"{o}dinger
equations and a new family of solvable models}
$$
$$
 {\bf V\'{\i}t Jakubsk\'{y}\footnote{jakub@ujf.cas.cz},}\\
{\small Nuclear Physics Institute, \v{R}e\v{z}, Czech Republic ,\\
 FNSPE, CTU, Prague, Czech Republic,}\\

 {\bf Miloslav Znojil\footnote{znojil@ujf.cas.cz},}\\
 {\small Nuclear Physics Institute, \v{R}e\v{z}, Czech Republic ,}\\

 {\bf Euclides Augusto
 Lu\'{\i}s\footnote{eluis@math.ist.utl.pt},}\\
  {\small CEMAT, IST, Lisboa, Portugal,}\\

 {\bf Frieder Kleefeld\footnote{kleefeld@gtae3.ist.utl.pt}}\\
 {\small CFIF, IST, Lisboa, Portugal}\\
\end{center}

\begin{abstract}
Angular parts of certain solvable models are studied. We find that
an extension of this class may be based on suitable trigonometric
identities. The new exactly solvable Hamiltonians are shown to
describe interesting two- and three-particle systems of the
generalized Calogero, Wolfes and Winternitz-Smorodinsky types.

\end{abstract}

\section{Introduction}
A family of superintegrable models has been introduced
in~\cite{winter}. One of them describes the movement of a particle
in $D$-dimensional space and has the following Hamiltonian
 \be H=\sum_{k=1}^{D}\left[-\frac{\partial^2}{\partial
 x_k^2}+\frac{\omega^2}{2}x_k^2+\frac{g_k(g_k-1)}{x_k^2}\right]\label{SW}.
 \ee
The domain of definition is the set of functions which belong
together with their first and second derivatives to
$L^2(\bigotimes_{k=1}^D[(-\infty,0)\cup(0,\infty)])$ and which
vanish at $x_i=0,\ i=1,..,D$. Due to its simplicity,
Hamiltonian~(\ref{SW}) serves as a useful playground for various
methods in Quantum Mechanics. For $D=2$, the pertaining
Schr\"{o}dinger equation is separable not only in cartesian but
also in polar and elliptical coordinates. In polar
coordinates~($x_1=r\cos\phi,\ x_2=r\sin\phi$), the angular
equation is
 \be \Omega\xi_l(\phi)=\left[-\frac{\partial^2}{\partial \phi^2}+\frac{g_1(g_1-1)}{\sin^2\phi}+
 \frac{g_2(g_2-1)}{\cos^2\phi}\right]\xi_l(\phi)=b_l^2\xi(\phi)\label{PeschlTeller},\ee
where the solutions coincide with the well known
P\"oschl-Teller~\cite{flugge} states defined in terms of Jacobi
polynomials,
 \be
 \xi^{(g_1,g_2)}_n(\phi)\sim\sin^{g_1}\phi\cos^{g_2}\phi
 P_n^{(g_1-\frac{1}{2},g_2-\frac{1}{2})}(\cos 2\phi),
 \label{solPeschlTeller}
 \ee
 $$b_l^2=(g_1+g_2+2n)^2.$$
In the most elementary special case with $g_1=g$ and $g_2= 0$ the
domain of definition of our solutions is a union of two subdomains
in the $x_1 - x_2$ plane, separated by an impenetrable barrier. In
the language of $\phi$, we have to consider a union of two
intervals,
$Dom(\phi)=\bigcup_{k=0}^{1}\left({k\pi},{(k+1)\pi}\right)$.

Of course, a much more interesting model will work with both the
couplings $g_1$ and $g_2$ different from zero where, at equal
strengths $g_1=g_2=g\neq 0$, the wave functions are Gegenbauer
polynomials,
 \be \xi_n^{(g,g)}(\phi)\sim\sin^g2\phi C_n^{g}(\cos 2\phi)
 \label{gegenbauer}
 \ee
and coincide with eigenfunctions of the modification $
-\frac{\partial^2}{\partial
 \phi^2}+\frac{4g(g-1)}{\sin^22\phi}$ of the operator $\Omega$
in~(\ref{PeschlTeller}). This coincidence is the consequence of
the following simple trigonometric identity
 \be
 \frac{1}{\sin^2\phi}+\frac{1}{\cos^2\phi}=\frac{4}{\sin^22\phi}.\label{WTrigIdent}
 \ee
The simplicity of the latter identity looks indicative and
enigmatic at the same time. Firstly, it seems to reflect the
separability of our problem as well as a symmetry in positions of
the singularities on the circular domain, i.e., after some
elementary trigonometry,
 \be
 \frac{1}{\sin^2\phi}+\frac{1}{\sin^2(\phi-\pi/2)}
 +\frac{1}{\sin^2(\phi-\pi)}+\frac{1}{\sin^2(\phi-3\pi/2)}
 =\frac{8}{\sin^2 2\phi}.\label{WtygIdent}
 \ee
Secondly, the exceptionality of our choice of the identical
strengths $g$ might open an immediate relationship between the
model (\ref{SW}) and several other solvable models based on the
use of a suitable Lie algebra of symmetries
\cite{OlshanetskyPerelomov} (cf. also section 3.2 below and/or a
very recent developments as sampled in
refs.~\cite{turbiner1},~\cite{turbiner2} and~\cite{turbiner3}).
Finally, our interest in the elementary trigonometry proved
further enhanced by the recent independent clarification of the
solvability of certain models using q-deformed Coxeter groups
\cite{fring}.

An overlap of all these observations formed a motivation of our
forthcoming considerations.

\section{Auxiliary trigonometric identities}

It might be possible to find an identity resembling
(\ref{WTrigIdent}) when $\sin^{-2}$-type singularities cut the
circular domain into $N$ equal parts with an arbitrary integer
$N$,
$Dom(\phi)=\bigcup_{k=0}^{N-1}\left(\frac{2k\pi}{N},\frac{2(k+1)\pi}{N}\right)$.
Such a desired generalization has been found to possess the form
 \be\sum_{k=0}^{N-1}\frac{1}{\sin^2\left(\phi-\frac{2k\pi}{N}\right)}=
 \left\{\begin{array}{c} {\frac{N^2}{\sin^2 N\phi},\ N \ odd}
 \\{\frac{N^2}{2\sin^2 \frac{N}{2}\phi},\ N\ even}. \end{array}\right.
 \label{trig}\ee
The rigorous proof of its validity is both simple and
straightforward. We start from
 \be\sum_{k=0}^{N-1}\frac{1}{\sin^2\left(\phi-\frac{2\pi k}{N}\right)}
 =-\frac{d^2}{d\phi^2}\ln \left(\prod_{k=0}^{N-1}
 \sin\left(\phi-\frac{2\pi k}{N}\right)\right) \ee
and employ the known formula for the product of trigonometric
functions~\cite{Gradstein},
 \be\prod_{k=0}^{N-1}\sin\left(\phi-\frac{2\pi k}{N}\right)
 =\left\{\begin{array}{c}{\frac{(-1)^{\frac{1-N}{2}}}{2^{N-1}}\sin
 N\phi},\ N\ odd
 \\{\
 \frac{(-1)^{\frac{-N}{2}}}{2^{N-1}}(1-\cos N\phi)},\ N\ even\end{array}\right.
 \ee
This gives eq. (\ref{trig}) immediately. A cosine analog
to~(\ref{trig})
 \be\sum_{k=0}^{N-1}\frac{1}{\cos^2\left(\phi-\frac{2k\pi}{N}\right)}=
 \left\{\begin{array}[l]{c} {\frac{N^2}{\cos^2 N\phi},\ N=2p+1}
 \\{\frac{N^2}{2\cos^2 \frac{N}{2}\phi},\ N=4p+2}
 \\{\frac{N^2}{2\sin^2 \frac{N}{2}\phi},\ N=4p, \ \ \ p\in\mathbb{N}}. \end{array}\right.
 \label{trigbe}\ee
can be proved in the similar manner. We have to keep in mind that
for the real arguments $\phi$, the identities contain the periodic
trigonometric functions. Even if we move to the complex arguments
$\phi$~\cite{znojil1},~\cite{znojil2},~\cite{ja}, the emergence of
the phase of $\phi$ merely introduces a new rather artificial
``degree of freedom" while the trigonometric identities themselves
remain unchanged.

For the real $\phi$ we are now prepared to
regard~(\ref{PeschlTeller}) as an  $N=1$ special case of a much
broader class of the exactly solvable angular Schr\"odinger
equations $\Omega_{(N)}\xi_m(\phi)=b_m^2\xi_m(\phi)$, i.e.,
 \be
    \left(-\frac{\partial^2}{\partial \phi^2}+
 \sum_{k=0}^{N-1}\frac{g_1(g_1-1)}{\sin^2\left(\phi-\frac{2k\pi}{N}\right)}+
 \sum_{l=0}^{N-1}\frac{g_2(g_2-1)}{\cos^2\left(\phi-\frac{2l\pi}{N}\right)}
 \right)\xi_m(\phi)
 =b_m^2\xi_m(\phi)\label{genPoeschlTeller}\ee
The eigenvalues are given by the following formulas for
corresponding integers $N$
 \be b_m=\frac{N}{4}\left(1+\sqrt{1+8[g_1(g_1-1)+g_2(g_2-1)]}+4m\right), \ \ \ N=4p\nonumber\ee
 \be b_m=\frac{N}{4}\left(2+\sqrt{1+8[g_1(g_1-1)]}+\sqrt{1+8[g_2(g_2-1)]}+4m\right),\ \ \ N=4p+2\nonumber\ee
 \be b_m=N(g_1+g_2+2m),\ \ N=2p+1,\  p,m\in\mathbb{N}.\label{energy}\ee
Also the solvability of~(\ref{genPoeschlTeller}) in terms of
Jacobi polynomials is retained since we can perform the summations
in the angular part $V$ of the interaction,
 \bea V&=&\frac{N^2}{2}\left[\frac{g_1(g_1-1)+g_2(g_2-1)}{\sin^2\frac{N}{2}\phi}\right]
 ,\ \ \ N=4p\nonumber\\
 &=&\frac{N^2}{2}\left[\frac{g_1(g_1-1)}{\sin^2\frac{N}{2}\phi}
 +\frac{g_2(g_2-1)}{\cos^2\frac{N}{2}\phi}\right],\ \ \ N=4p+2\nonumber\\
 &=&N^2\left[\frac{g_1(g_1-1)}{\sin^2N\phi}+\frac{g_2(g_2-1)}{\cos^2N\phi}\right],
 \ \ \ N=2p+1,\ p\in\mathbb{N}
 \label{solvability}
 \eea
using the identities~(\ref{trig}) and~(\ref{trigbe}).

\section{A few immediate applications}

We may return back from eq. (\ref{solvability}) to the
P\"oschl-Teller bound-state problem in one dimension after an
elementary linear transformation $\tilde{\phi}=N\phi$ or
$\tilde{\phi}=\frac{N}{2}\phi$. Similarly, some of the related
known separable and solvable models in more dimensions may be
revealed as special cases as well. Nevertheless, our present key
message is that in the latter context, also some new solvable
models emerge due to our full freedom in the choice of the integer
$N$ in~(\ref{genPoeschlTeller}).

\subsection{One- and two-particle context}

Once we leave the radial part of the (separable) partial
differential eq.~(\ref{SW}) unchanged, our task is to perform just
a backward transition to the original, ``physical" cartesian
coordinates $y_1=r\cos\phi,\ y_2=r\sin\phi$. Introducing the fixed
constant parameters $s_k=\sin\frac{2k\pi}{N}$ and
$c_k=\cos\frac{2k\pi}{N}$ we get the new form of the Hamiltonian,
 \be
  H=-\frac{\partial^2}{\partial y_1^2}-\frac{\partial^2}{\partial y_2^2}
 +\frac{\omega^2}{2}\left(y_1^2+y_2^2\right)+\sum_{k=0}^{N-1}
 \frac{g_2(g_2-1)}{\left(y_1c_k+y_2s_k\right)^2}+
 \sum_{l=0}^{N-1}\frac{g_1(g_1-1)}{\left(y_1s_l-y_2c_l\right)^2}, \label{genBC2}
 \ee
where the mathematical separability of our general physical
bound-state model remains ``hidden". Its energies are easily
expressible in the form
$E_{n,m}=\sqrt{2}\omega(2n+\frac{\sqrt{2}}{2}b_m+1)$ where the
eigenvalues $b_m$ of the angular equation itself are written
explicitly in~(\ref{energy}).

The latter formula may be read as characterizing a new and rather
general two particle model where the interaction acquires the
different forms for the different values of the integer $N$. In
this way the most elementary choice of $N=1$ returns us back to
the Smorodinsky-Winternitz model~(\ref{SW}).

At $N=2$ the interaction looks much more complicated. Fortunately,
after its brief inspection we reveal that it coincides with the so
called $BC_2$ model of the exactly solvable
type~\cite{OlshanetskyPerelomov},
 \be H_{BC_2}=\sum_{i=1}^2\left[-\frac{\partial^2}{\partial y_i^2}
 +\frac{\omega^2}{2}y_i^2+\frac{g_1(g_1-1)}{y_i^2}\right]
 +\left(\frac{g_2(g_2-1)}{(y_2-y_1)^2}+\frac{g_2(g_2-1)}
 {(y_2+y_1)^2}\right).\label{bc_2}\ee
In this sense we can consider~(\ref{genBC2}) as a common
generalization of the two mathematically different and apparently
physically uncorrelated models.

\subsection{Three-particle setting}

In the spirit of what has been said in Introduction,
Hamiltonians~(\ref{genBC2}) can easily be re-interpreted as
describing {\em three} interacting particles on the line. It
suffices to consider $y_i$ as the two, so called Jacobi
coordinates of the system, to be complemented by the third, so
called center-of-mass (CMS) coordinate of the whole triplet.
Formally, the transformation of the physical single-particle
cartesian coordinates $x_1,\ x_2,\ x_3$ into the CMS ones is given
by the well known formula,
 \be\left(\begin{array}{c}y_1\\y_2\\Y\end{array}\right)=
 \left(\begin{array}{ccc}-\frac{\sqrt{2}}{2}&\frac{\sqrt{2}}{2}&0\\
 \frac{\sqrt{6}}{6}&\frac{\sqrt{6}}{6}&-\frac{\sqrt{6}}{6}\\
 \frac{\sqrt{3}}{3}&\frac{\sqrt{3}}{3}
 &\frac{\sqrt{3}}{3}
 \end{array}\right)\left(\begin{array}{c}x_1\\x_2\\x_3\end{array}\right),\label{CMS}\ee
where $Y$ is the coordinate of the center of mass. We add
to~(\ref{genBC2}) a kinetic and potential term
$-\frac{\partial^2}{\partial Y^2}+\frac{\omega^2}{2}Y^2$
responsible, as usual, for the confined oscillatory motion of the
center of mass of the whole system.

Using a transformation which is inverse to~(\ref{CMS}) we are now
getting a new and interesting model of particles which interact
via both an attractive and repulsive one-, two- and three-particle
interaction in general. The new Hamiltonian reads
 \bea H&=&-\frac{1}{2}\frac{\partial^2}{\partial x_1^2}-\frac{1}{2}\frac{\partial^2}{\partial
 x_2^2}-\frac{1}{2}\frac{\partial^2}{\partial x_3^2}+
 \frac{1}{2}\omega^2 \left(x_1^2+x_2^2+x_3^2\right)\nonumber\\
 &&+
 \sum_{k=0}^{N-1}\frac{2g_2(g_2-1)}{\left[\left(c_k+\frac{\sqrt{3}}{3}s_k\right)x_1
 +\left(-c_k+\frac{\sqrt{3}}{3}s_k\right) x_2-\frac{2\sqrt{3}}{3}s_kx_3\right]^2}
 \nonumber\\ && +
 \sum_{l=0}^{N-1}\frac{2g_1(g_1-1)}{\left[\left(-s_l+\frac{\sqrt{3}}{3}c_l\right)x_1
 +\left(s_l+\frac{\sqrt{3}}{3}c_l\right)x_2
 -\frac{2\sqrt{3}}{3}c_lx_3\right]^2}\label{genG2}\eea
and its spectrum is purely discrete. The explicit form of the
energies
 \be
 E_{n,m,t}=\sqrt{2}\omega(2n+\frac{\sqrt{2}}{2}b_m+1)+\sqrt{2}\omega(t+\frac{1}{2})
 \ \ n,m,t\in \mathbb{N}
 \label{energy 3 body}
 \ee
contains a  term which emerges due to an overall confinement of
the whole triplet in the harmonic-oscillator well (= the contained
movement of the center of the mass) and the term which depends
directly on the angular-equation eigenvalue $b_l$
(cf.~(\ref{energy})).

\section{Discussion}

We may summarize that in the context of section 3.1 our new
solvable Hamiltonians describe a particle which moves over a
complicated potential surface containing also some strongly
singular barriers. In the parallel angular-equation
re-interpretation of our trigonometric identities (and their
consequences) in section 3.2 we arrived at a new family of the
genuine three particle models which remain exactly solvable but
which remain solely separable in the ``unphysical", auxiliary
polar coordinates.

A few more comments may be added.

\subsection{The variability of $N$}

Similarly to the generalized two-particle system~(\ref{genBC2}),
the present Hamiltonian (\ref{genG2}) acquires the form of a well
known system at $N=3$ and at the very special coupling strengths.

For $g_2=0$, eq.~(\ref{genG2}) describes the well-known
three-particle Calogero model~\cite{calogero}, whose potential is
 \be V_{Cal}=\frac{\omega^2}{2}\left(x_1^2+x_2^2+x_3^2\right)+\sum_{j<k}^3\frac{g_1(g_1-1)}
 {(x_k-x_j)^2}.\nonumber\ee
If we let $0\neq g_1\neq g_2\neq 0$, the  Wolfes
model~\cite{Wolfes} with the following potential is revealed,
 \be V_{W}=\frac{\omega^2}{2}(x_1^2+x_2^2+x_3^2)+\sum_{j<k}^3\frac{g_1(g_1-1)}{(x_k-x_j)^2}
 +\sum_{l<m,l,m\neq n}^3\frac{g_2(g_2-1)}{(x_l+x_m-2x_n)^2}.
 \ee
In the next step let us evaluate the repulsive potential
of~(\ref{genG2}) for a few higher values of the index $N$.

At $N=8$, formula~(\ref{solvability}) suggests that the couplings
will merge
and form only one type of singular potential with coupling strength $g=(g_1(g_1-1)+g_2(g_2-1))$\\
 \bea V_{N=8}&=&g\left(\frac{4}{(x_1-x_2)^2}+\frac{12}{(x_1+x_2-2x_3)^2}\right.\nonumber\\
 &&\left.+\sum_{\varepsilon=+,-}\frac{2}{\left[\left(\varepsilon\frac{1}{2}+\frac{\sqrt{3}}{3}\right)x_1
 +\left(-\varepsilon\frac{1}{2}+\frac{\sqrt{3}}{3}\right)x_2-\frac{\sqrt{3}}{3}x_3\right]^2}
 \right).\eea
A very different situation occurs at $N=5$ where there emerge
singularities of two types distinguished by the coupling strengths
$g_1$ and $g_2$,
 \bea V_{N=5}&=&\sum_{k=1,2}\sum_{\varepsilon=\pm}\left[\frac{2g_2(g_2-1)}
 {\left[\left(c_k+\varepsilon
 \frac{\sqrt{3}}{3}s_k\right)x_1+
 \left(-c_k+\varepsilon \frac{\sqrt{3}}{3}s_k\right)x_2 -\varepsilon
 \frac{2\sqrt{3}}{3}s_kx_3\right]^2}
 \right]\nonumber\\
 &&\left.\frac{2g_1(g_1-1)}{\left[\left(\frac{\sqrt{3}}{3}c_k-\varepsilon
 s_k\right)x_1+
 \left(\frac{\sqrt{3}}{3}c_k+\varepsilon s_k\right)x_2 -
 \frac{2\sqrt{3}}{3}c_kx_3\right]^2}
 \right]\nonumber\\
 &&+\frac{2g_2(g_2-1)}{(x_2-x_1)^2}+\frac{6g_1(g_1-1)}{(x_1+x_2-2x_3)^2}\eea
where one only has to keep in mind that $c_k=\cos\frac{k\pi}{5}$
and $s_k=\sin\frac{k\pi}{5}$.

\subsection{ Outlook}

Being inspired by the simple superintegrable model~(\ref{SW}), we
derived the class of the trigonometric identities~(\ref{trig})
and~(\ref{trigbe}) by the entirely elementary mathematical means.
These identities proved to be an unexpectedly productive tool for
a generalization of several known solvable models. At the same
time, the emergence and solvability of the  new Hamiltonians opens
many new questions.

We did not manage to touch many of them in this text. First of
all, one must ask whether there exists an alternative or deeper
algebraic background of their solvability. Next, even on a purely
analytic level of our considerations a deeper insight would be
welcome concerning the role of the repulsive barriers. Last but
not least, a guidance towards a future analysis of the models with
multiple barriers might be also sought somewhere in between their
single-particle and multi-particle special cases. Thus, one would
really appreciate seeing {\em more} formal parallels between the
superintegrable, {\em separable} Winternitz-Smorodinsky-type
systems {in more dimensions} and the intrinsically nonseparable
{\em more-particle} systems belonging to the algebraically
classified Calogero-type family.

\section*{Acknowledgement}
This work was supported by AS CR grant No. A 1048302, and by the
Funda\c{c}\~{a}o para a Ci\^{e}ncia e a Tecnologia (FCT) of the
{\em Minist\'{e}rio da Ci\^{e}ncia e do Ensinio Superior} of
Portugal, under grant No. SFRH/BDP/9480/2002.

\end{document}